\documentstyle[aps,prd,preprint,eqsecnum,epsf]{revtex}
\input epsf.tex

\newcommand{\be}{\begin{equation}}
\newcommand{\ee}{\end{equation}}
\newcommand{\ba}{\begin{eqnarray}}
\newcommand{\ea}{\end{eqnarray}}
\newcommand{\bas}{\begin{eqnarray*}}
\newcommand{\eas}{\end{eqnarray*}}
\newcommand{\hsp}{\hspace{.5cm}}

\newcommand{\spaceand}{\hsp {\rm and} \hsp}

\tightenlines

\begin{document}

\draft

\title{Rotation and the AdS/CFT correspondence}
\author{S.W. Hawking\thanks{email:  swh1@damtp.cam.ac.uk}, 
        C.J. Hunter\thanks{email:  C.J.Hunter@damtp.cam.ac.uk} and 
M. M. Taylor-Robinson\thanks{email: M.M.Taylor-Robinson@damtp.cam.ac.uk}}
\address{Department of Applied Mathematics and
      Theoretical Physics, University of Cambridge,
      \\Silver Street, Cambridge CB3 9EW, United Kingdom }
\date{\today}

\maketitle

\begin{abstract}
{
In asymptotically flat space a rotating black hole cannot be 
in thermodynamic equilibrium because the thermal radiation would 
have to be co-rotating faster than light far from the black hole. 
However in asymptotically anti-de Sitter space such equilibrium is 
possible for certain ranges of the parameters. We examine the relationship
between conformal field theory in rotating Einstein universes of 
dimensions two to four and Kerr anti-de Sitter black holes in dimensions 
three to five. The five dimensional solution is new. We find similar 
divergences in the partition function of the conformal field theory 
and the action of the black hole at the critical angular velocity 
at which the Einstein rotates at the speed of light. This should be 
an interesting limit in which to study large $N$ Yang-Mills. 
}

\end{abstract}



\section{Introduction}
\label{sec:intro}

In Minkowski space the only Killing vector that is time like
everywhere is the time translation Killing vector 
$\partial /\partial t $. For instance, in four dimensional Minkowski
space, the Killing vector 
$\partial /\partial t + \Omega \partial /\partial \phi $ that
describes a frame rotating with angular velocity $\Omega $ becomes 
space like outside the velocity of light cylinder 
$r\sin \theta =1/\Omega $. 

This raises problems with the 
thermodynamic interpretation of the Kerr solution: a Kerr solution 
with non zero rotation parameter $a$ cannot be in equilibrium with 
thermal radiation in infinite space because the radiation would have 
to co-rotate with the black hole and so would have to move faster 
than light outside the velocity of light cylinder. The best one can 
do is consider the rather artificial case of equilibrium with 
rotating radiation in a box smaller than the velocity of light
radius. This problem is inextricably linked with
the fact that the Hartle-Hawking state for a Kerr solution does not
exist, as proved in \cite{Ka_Wa}. The absence of the Hartle-Hawking
state has a number of important ramifications, details of which are
discussed in \cite{Ka_Wa}.  

On the other hand, even a non rotating Schwarzschild black 
hole has to be placed in a finite sized box because otherwise the 
thermal radiation would have infinite energy and would collapse 
on itself. There is also the problem that the equilibrium is 
unstable because the specific heat is negative.

\bigskip

It is now well known \cite{HP}, \cite{Wit} that the specific heat of large 
Schwarzschild anti de Sitter black holes is positive and that the red 
shift in anti-de Sitter spaces acts like an effective box to remove the 
infinite energy problem. What was less well known except in the rather
special three dimension case was that anti-de Sitter boundary
conditions could also remove the faster than light problem 
for rotating black holes. That is, in anti-de Sitter space there are 
Killing vectors that are rotating with respect to the standard time
translation Killing vector and yet are timelike everywhere. This
means that one can have rotating black holes that are in equilibrium 
with rotating thermal radiation all the way out to infinity.

One would expect \cite{M;G;W}, \cite{Wit} the partition function of this 
black hole to be related to the partition function of a conformal
field theory in a rotating Einstein universe on the boundary of 
the anti-de Sitter space. It is the aim of this paper to examine 
this relationship and draw some surprising conclusions.

\bigskip

Of particular interest is the behaviour in the limiting case in 
which rotational velocity in the Einstein universe at infinity a
approaches the speed of light. We find that the actions of the 
Kerr-AdS solutions in four and five dimensions 
have similar divergences at the critical angular velocity to the partition 
functions of conformal field theories in rotating Einstein universes of one 
dimension lower. This is like the behaviour of the three dimensional  
rotating anti-de Sitter black holes and the corresponding 
conformal field theory on the two dimensional Einstein universe or cylinder. 
There is however an important 
difference: in three dimensions one calculates the actions of the BTZ black 
holes relative to a reference background that is the $M=0$ BTZ black
hole. Had one used three dimensional anti-de Sitter space as the 
reference background, one would have had an extra term in the action 
which would have diverged as the critical angular velocity was
reached. 

On the conformal theory side, this choice of reference
background is reflected in a freedom to choose the vacuum energy. 
However, in higher dimensions there is no analogue of the $M=0$ BTZ 
black hole to use as a reference background. One therefore has to use 
anti-de Sitter space itself as the reference background. Similarly, 
there isn't a freedom to choose the vacuum energy in the conformal 
field theory. Any mismatch between the reference background for 
anti-de Sitter black holes and the vacuum energy of the conformal 
field theory will become unimportant in the high temperature limit 
for non rotating black holes or the finite temperature but critical 
angular velocity case. Thus it might be that the black hole/thermal 
conformal field theory correspondence is valid only in those 
limits. In that case, maybe we shouldn't believe that the large 
$N$ Yang Mills theory in the Einstein universe has a phase transition.

\bigskip

In the $1+1$ dimensional boundary of three dimensional anti-de Sitter space, 
massless particles move to the left or right at the speed of light. The 
critical angular velocity corresponds to all the particles moving in the same 
direction. If the temperature is scaled to zero as the angular velocity 
approaches its critical value, the energy remains finite and the system 
approaches a BPS state.

In higher dimensional Einstein universes however particles can move in 
transverse directions as well as in the rotation direction or its opposite. At 
zero angular velocity, the velocity distribution of thermal particles is 
isotropic but as the angular velocity is increased the velocity distribution 
becomes peaked in the rotation direction. When the rotational velocity reaches 
the speed of light, the particles would have to be moving exclusively in the 
rotation direction. This is impossible for particles of finite energy. Thus 
rotating Einstein universes of dimension greater than two cannot
approach a finite energy BPS state as the
angular velocity approaches the critical value for rotation at the 
speed of light. 

Corresponding to this, we shall show that four and 
five dimensional Kerr-AdS solutions do not approach a BPS 
state as the angular velocity approaches the critical value,
unlike the three dimensional BTZ black hole. Nevertheless critical 
angular velocity may be of interest because one might expect that 
in this limit super Yang-Mills would behave like a free theory. We 
postpone to a further paper the question of whether this removes 
the apparent discrepancy between the gravitational and Yang Mills 
entropies.

We should mention that
critical limits on rotation have recently been discussed in the
context of black three branes in type IIB supergravity \cite{gubs}:
rotating branes are found to be stable only up to a critical value of
the angular momentum density, beyond which the specific heat becomes
negative. However, our critical limit is different. It
corresponds not to a thermodynamic
instability, but rather to a Bose condensation effect in the boundary
conformal field theory. 

\bigskip

In section two we calculate the partition function for conformal 
invariant free fields in rotating Einstein universes of dimension two,
three and four in the critical angular velocity limit. In sections 
three, four and five we calculate the entropy and actions for 
rotating anti-de Sitter black holes in the
corresponding dimensions and find agreement with the conformal 
field in the behaviour near the critical angular velocity. 

The metric for 
rotating anti-de Sitter black holes 
in dimensions higher than four was not previously known. Our solutions
have other interesting applications, particularly when regarded as
solutions of gauge supergravity in five dimensions, which we will
discuss elsewhere \cite{mmt}. 

\section{Conformally invariant fields in rotating Einstein universes}

The Maldacena conjecture \cite{M;G;W}, \cite{Wit}
implies that the thermodynamics of quantum gravity
with a negative cosmological constant can be modelled by the large $N$
thermodynamics of quantum field theory. 
We are interested here in probing the
correspondence in the limit that the boundary is rotating at the speed
of light; that is, we want to study the large $N$ thermodynamics of
conformal field theories in an Einstein universe rotating at the speed
of light. 

The details of the boundary conformal field theory ultimately 
depend on the details of the bulk supergravity (or string) theory, but
generic features such as the divergence of the entropy in this
critical limit should be independent of the precise features of the
theory. Thus we are led to making the following simplification:
instead of considering, for example, the large $N$ limit of ${\cal
  N}=4$ SYM in four dimensions we can just look at the behaviour of
conformal scalar fields in a rotating Einstein universe. We find that
this does indeed give us generic thermodynamic features at high
temperature which agree with those found from the bulk theory. 

To go further than this, we would have to embed the rotating black hole
solutions within a theory for which we know the corresponding
conformal field theory. For instance, we could embed the five
dimensional anti-de Sitter Kerr black holes into IIB supergravity in
ten dimensions; we then know that the corresponding conformal field
theory is the large $N$ limit of ${\cal N}=4$ SYM. 
However, since we can't calculate
quantities in the large $N$ limit of the latter, to obtain the
subleading behaviour of the partition function would require some
approximations or models such as those used in the discussion of
rotating three branes in \cite{gubs}. It would be interesting to 
show that the perturbative SYM calculation gives a discrepancy of
$4/3$ in the entropy as one expects from the results of
\cite{Gu_Kl_Pe}.   

Of course in two dimensions we can do better than this:
the two-dimensional conformal field theory is well understood in the
context of an old framework \cite{Br_He}, where the correspond between
bulk and boundary is effectively provided by the modular invariance of
the boundary conformal field theory \cite{Cardy}, \cite{Strominger97}.
In recent months, the CFT
has been discussed in some detail, for example in \cite{Ma_St}, and one
should be able to obtain the subleading dependences of the partition
function on the angular velocity $\Omega$. We leave this 
issue to future work. 

It is interesting to note here that there is no equivalent of the zero mass
BTZ black hole in higher dimensions. Since the correspondence between
the bulk theory and the boundary conformal field theory is clearest
when one takes the background to be the BTZ black hole, the
correspondence between the conformal field theory and supergravity in
the anti-de Sitter background may only be approximate in higher
dimensions, valid for high temperature. This is one reason why it is
useful to investigate what happens in the critical angular
velocity limit. 

\bigskip

Let us start with an analysis of conformal fields in a two-dimensional
rotating Einstein universe; the metric on a cylinder is 
\be
ds^2 = - dT^2 + d\Phi^2,
\ee
where we need to identify $\Phi \sim \Phi + \beta \Omega$, and both
the inverse temperature $\beta$ and the angular velocity $\Omega$ are
dimensionless. Now consider modes of a conformally coupled scalar
field, propagating in this background; for harmonic modes, the
frequency $\omega$ is equal in magnitude to the angular momentum
quantum number $L$. So we can write the partition function for
conformally invariant scalar fields as
\be
\ln {\cal Z} = - \sum \ln \left ( 1 - e^{-\beta (\omega - L
    \Omega)} \right ) - \sum \ln \left (1 - e^{-\beta (\omega + L
    \Omega)} \right ),
\ee
where the first term counts left moving modes and the second term
counts right moving modes. The partition
function is manifestly singular as one takes the limit $\Omega
\rightarrow \pm 1$; in this limit, all the particles rotate in one
direction. Provided that $\beta$ is small we 
can approximate the summation by an integral so that 
\be 
\ln {\cal{Z}} \approx \frac{\pi^2}{6 \beta (1-\Omega^2)},
\ee
which agrees with the high temperature result found in the next
section (\ref{dim_act}) 
up to a factor and a scale $l$. Note that the form of this result
could also be derived by requiring conformal invariance in the high
temperature limit. 

\bigskip

Let us now consider the conformal field theory in three dimensions;
a hypersurface of constant large radius in the four-dimensional 
anti-de Sitter Kerr metric has a metric which is proportional to 
a three dimensional Einstein universe
\be
ds^2 = -dT^2 + d\Theta^2 + \sin^2\Theta d\Phi^2,
\ee
where $\Phi$ must be identified modulo $\beta
\Omega$ with $\beta$ and $\Omega$ dimensionless. 
Now consider a conformally
coupled scalar field propagating in this background: the field
equation for a harmonic scalar is  
\be
\left (\nabla - \frac{R_g}{8} \right ) = 
\left ( \nabla - \frac{1}{4} \right ) \varphi = 0,
\ee
where $\nabla $ is the d'Alambertian and $R_{g}$ is the Ricci scalar. 
Modes of frequency $\omega$ satisfy the constraint 
\be
\omega^2 = L(L+1) + \frac{1}{4} = (L + \frac{1}{2})^2,
\ee
where $L$ is the angular momentum quantum number. Then the partition
function can be written as 
\be 
\ln {\cal Z} = - \sum_{L=0}^{\infty} \sum_{m=-L}^{L} 
\ln \left ( 1 - e^{-\beta(\omega - m\Omega)}
\right). \label{par_fun}
\ee
For small $\beta$ we can approximate this summation as the integral 
\be
\ln {\cal Z} \approx - \int_{0}^{\infty} dx_{L} \int_{-x_{L}}^{x_{L}}
  dx_{M} \ln \left ( 1 - e^{-\beta(x_{L} -\Omega x_{M})} 
\right )= \frac{1}{\beta^2} \int_{0}^{\infty} dy \int_{-y}^{y} dx 
\ln \left ( 1 - e^{-(y -\Omega x)}  \right ) \label{sum_fun}
\ee
We are interested in the divergence of the partition function when
$\Omega \rightarrow \pm 1 $; this divergence arises from the modes for
which the frequency is almost equal to $\left | m \right | $. 
Of course the frequency can never be quite equal to 
$\left | m \right | $, 
but for large $m$ the argument of the logarithm in (\ref{par_fun}) 
becomes very small. So picking out the modes for which $y= \left | x
  \right | $ in
(\ref{sum_fun}) we find that the leading order divergence in the
partition function at small $\beta$ is 
\be
\ln {\cal Z} \approx \frac{\pi^2}{6 \beta^2 (1 - \Omega^2)},
\ee
which agrees in functional form with the limit that we will find for the
bulk action in section four. In the critical limit, all the particles
are rotating at the speed of light in the equatorial plane. 

\bigskip

The metric of the four-dimensional rotating Einstein universe can be
written as  
\be
ds^2 = - dT^2 + d\Theta^2 + \sin^2\Theta d\Phi^2 + \cos^2\Theta
d\Psi^2,
\ee
where $\Phi$ and $\Psi$ must be identified modulo $\beta \Omega_1$ and
$\beta \Omega_2$. We have only approximated the partition function 
for conformally coupled scalar fields in lower dimensional 
rotating Einstein universes. However in \cite{CassidyHawking97} the 
thermodynamics of conformally coupled scalars were discussed in detail 
for a four-dimensional rotating Einstein universe in the limit in
which one of the angular velocities vanishes.  
The general form for the 
partition function found in \cite{CassidyHawking97} is quite
complex, but it takes a 
simple form when $\beta$ is small: one finds that 
\be
\ln {\cal Z} \approx \frac{\pi^3}{90 \beta^3 (1 - \Omega^2)}, \label{mjc}
\ee
where $\Omega$ is the angular velocity, which agrees in form with the
bulk result to leading order. In principle we could use the partition
function density given in \cite{CassidyHawking97} to probe the
correspondence between subleading terms. 

Let us now try to approximate the partition function for general
angular velocities using the same techniques as before. 
Consider a conformally invariant scalar field propagating in this
background; the field equation is 
\be
\left ( \nabla - \frac{R_g}{6} \right ) = 
\left ( \nabla - 1 \right )\varphi = 0,
\ee
and so modes of the field have frequencies $\omega$ which satisfy
\be 
\omega^2 = L(L+2) + 1 = (L+1)^2,
\ee
where $L$ is the orbital angular momentum number. Then the partition
function may be written as 
\be
\ln {\cal Z} = - \sum_{L,m_1,m_2} \ln \left ( 1 - e^{-\beta (\omega - m_1
    \Omega_1 - m_2 \Omega_2)} \right),
\ee
where 
$m_1$ and $m_2$ are orbital quantum numbers. Suppose that $\Omega_2 = 0$;
then we expect the dominant contribution to the partition function in
the critical angular velocity limit to be from the $m_1 = \pm L$
modes. However there is a constraint on the angular momentum quantum
numbers   
\be
\left | m_1 \right | + \left | m_2 \right | \le L, \label{rot_con}
\ee
and so we need to set $m_2 = 0$. The dominant contribution to the
partition function at high temperature can be expressed as 
\be
\ln {\cal Z} \approx  - \frac{1}{\beta^3} \int_{0}^{\infty} dx \left [ \ln 
  \left ( 1 - e^{(1 + x (1- \Omega))} \right) + \ln 
  \left ( 1 - e^{(1 + x (1+ \Omega))} \right)  \right ] 
=  \frac{\pi^2}{6 \beta^3 (1-\Omega^2)}, \nonumber 
\ee
which agrees with the result (\ref{mjc}) in functional dependence
although not coefficient. 

\bigskip

For general angular velocities we find that the factor 
\be
\left (L - m_1 \Omega_1 - m_2 \Omega_2 \right ) 
\ee
only approaches zero in the limit $\Omega_1, \Omega_2 \rightarrow
1$. Thus we expect that there is a divergent contribution to the
partition function only when either or both of $\Omega_1$ and
$\Omega_2$ tend to
one, as we will find when we look at the black hole metric. 

Setting $\Omega_1 = \Omega_2 \equiv \Omega$, the dominant contribution to the
partition function will come 
from modes for which the bound (\ref{rot_con}) is saturated. Then
we find that 
\be
\ln {\cal Z}  \approx  - \frac{1}{\beta^3} 
\int_{0}^{\infty} dx x \left [\ln \left ( 1 - e^{-x(1-\Omega)} \right)
  + \ln \left ( 1 - e^{-x(1-\Omega)} \right) \right ] 
= \frac{\zeta(3)}{\beta^3 (1 - \Omega^2)^2},  
\ee
which has the correct dependence on $\beta$ and $\Omega$ to agree
with the bulk result found in section five. 

\section{Rotating black holes in three dimensions}
\subsection{The BTZ black hole}

The Euclidean Einstein action in three dimensions can be written as 
\be
I_{3} = - \frac{1}{16 \pi } \int d^3x \sqrt{g} \left [ R_g + 2 l^2
\right],
\ee
with the three dimensional Einstein constant set to one. The
Lorentzian section of the BTZ black
hole solution first discussed in \cite{BanadosTeitelboimZanelli92} is
\be 
ds^2 = - N^2 dT^2 + \rho^2 (N^{\Phi} dT + d\Phi)^2 +
(\frac{y}{\rho})^2 N^{-2} dy^2,
\ee
where the squared lapse $N^2$, the angular shift $N^{\phi}$ and the
angular metric $\rho^2$ are given by 
\ba
N^2 &=& (\frac{y l}{\rho})^2 ({y^2 - y_{+}^2}); \nonumber \\
N^{\Phi} = - \frac{j}{2 \rho^2}; && \hspace{5mm} \rho^2 = y^2 +
\frac{1}{2} (m l^{-2} - y_{+}^2),
\ea
with the position of the outer horizon defined by
\be
y_{+}^2 = m l^{-2} \sqrt{ 1 - (\frac{jl}{m})^2}.
\ee
Note that in these conventions anti-de Sitter spacetime is the $m=-1$,
$j=0$ solution. Cosmic censorship requires the existence of an event
horizon, which in turn requires either $m= -1$, $j=0$ or $m \ge \left
|j \right |l$. This bound in fact coincides with the supersymmetry
bound: regarded as a solution of the equations of motion of gauged
supergravity with zero gravitini, extreme black holes with $m = \left
| j \right | l$ have one exact supersymmetry. Both the $m=0$ and the
$m=-1$ black holes have two exact supersymmetries. In higher
dimensional anti-de Sitter Kerr black holes 
the cosmic censorship bound does not coincide with the
supersymmetry bound. 

The temperature of the black hole is given by 
\be
T_{H} = \frac{\sqrt{2m} l}{2 \pi} \left [ \frac{1 - (\frac{jl}{m})^2}{1 +
\sqrt{1 - (\frac{jl}{m})^2}} \right ]^{1/2}.
\ee
There has been a great deal of interest recently in the BTZ black
hole; the action was first calculated in
\cite{BanadosTeitelboimZanelli92} and has also been 
discussed in \cite{Ma_St}. However, the
action was calculated with respect to the zero mass black hole
background, whilst in the present context we are interested in the
action with respect to anti-de Sitter space itself. The reason for
this is that in higher dimensions there is no analogue of the zero
mass black hole as a background. {\footnote {The
   metric for which one replaces the lapse function $(1+ l^2 y^2)$ by 
 $l^2 y^2$ certainly plays a distinguished r\^{o}le in all dimensions,
 since this is the metric that one obtains from branes in the decoupling
 limit. It is not however true that this metric is the natural
 background for rotating black holes in dimensions higher than
 three but in the high temperature limit
 the distinction between the backgrounds will only affect subleading
 contributions to the action.}}

To calculate the action of the rotating black hole one first needs to
analytically continue both $t \rightarrow i \tau$ and $j \rightarrow
-i \bar{j}$. Using the Euclidean section one finds the action as a
function of $m$, $l$ and $\bar{j}$. The physical result is then
obtained by analytically continuing the angular momentum parameter. 
Taking the background to be anti-de Sitter space we then
find that the Euclidean action (for $m \ge 0$) is given by 
\be
I_3 = - \frac{\pi}{8 \sqrt{2m} l } \left [ \frac{1 + \sqrt{f}}{f}
\right ]^{\frac{1}{2}} \left [ 3m \sqrt{f} - (2 + m) \right ],
\ee
where $f = 1 - (jl/m)^2$. This action diverges in general 
as $f$ approaches zero, i.e. as we approach the cosmological and
supersymmetry bound. One would expect the action to diverge to
positive infinity in this limit; from the gravitational instanton point of
view, this implies that there is zero probability for anti-de Sitter
spacetime to decay into a supersymmetric BTZ black hole.  

It is straightforward to show that the energy $\cal{M}$, angular
momentum $J$, angular velocity $\Omega$ and entropy $S$ are given by
\ba
{\cal{M}} = \frac{1}{8} (m+1); && \hspace{5mm} J = \frac{j}{8};
\label{mass_btz} \\
S = \frac{1}{2} \pi \rho(y_{+}); && \hspace{5mm} \Omega = - \frac{j}{2
\rho^2(y_{+})}. \nonumber 
\ea
Note that the zero of energy is defined with respect to the 
anti-de Sitter space rather than the $m=0$ black hole.  

\bigskip

The asymptotic form of the Euclidean section of the BTZ metric is 
\be
ds^2 = y^2 l^2 d\tau^2 + y^2 d \Phi^2 + \frac{dy^2}{y^2 l^2}.
\ee
Regularity of the solution on the boundary of the Euclidean section 
at $y = y_{+}$ requires that 
we must identify $\tau \sim \tau + \beta$ and $\Phi \sim \Phi +
i \beta \Omega$, where $\beta$ is the inverse temperature. The latter
identification is necessary because the boundary is a fixed point set
of the Killing vector 
\be
k = \partial_{\tau} + i \Omega \partial_{\Phi}. 
\ee
The net result of these identifications is that after one analytically
continues back to Lorentzian signature one finds that 
the boundary at infinity is conformal to an
Einstein universe rotating at angular velocity $\Omega$. 

In the limit that $\Omega \rightarrow \pm l$ the surface is 
effectively rotating at
the speed of light: this gives the critical angular velocity 
limit. Looking back at the
form of the metric for the BTZ black hole we find that this limit
implies that  
\be 
\Omega = - \frac{jl^2}{m (1 + \sqrt{f})} \rightarrow \pm l,
\ee
which in turn requires that $f \rightarrow 0$. Hence in three
dimensions the cosmological and supersymmetry limits coincide with a
critical angular velocity limit. However, the temperature
necessarily vanishes whilst in the conformal field theory we have only
probed the high temperature limit. This suggests that one should be
able to find a more general critical angular velocity limit. This is
indeed the case: if we rewrite the BTZ metric in Kerr form we will be
able to find non-extreme states for which the boundary is rotating at
the speed of light. 

It is useful to rescale the time coordinate
so that $\hat{\beta}$ is both finite and dimensionless in the critical
limit
\be
\hat{\beta} = \sqrt{f} l \beta \approx \frac{2\pi}{\sqrt{2m}},
\ee
where the latter equality applies for $m$ large. 
In this limit of small $\hat{\beta}$ the action for the BTZ black hole
diverges as 
\be 
I_{3} \approx \frac{\pi^2}{8 l \hat{\beta} (1 - \hat{\Omega})},
\label{dim_act}
\ee
where $\hat{\Omega} = l^{-1} \Omega$ and is hence dimensionless. We
would need to know the CFT partition function at low temperature
to compare with the CFT and bulk results.

\subsection{Alternative metric for the BTZ black hole}
\noindent

To elucidate the thermodynamic properties of the black hole as one
takes the cosmological and supersymmetric limit
it is useful to rewrite the metric in the alternative form 
\be
ds^2 = - \frac{\Delta_{r}}{r^2} (dt - \frac{a}{\Xi} d\phi)^2 + \frac{r^2
dr^2}{\Delta_r} + \frac{1}{r^2} (a dt - \frac{1}{\Xi}(r^2 + a^2)
d\phi)^2, \label{alt_met}
\ee
where we define
\be
\Delta_r = (r^2 + a^2)(1 + l^2 r^2) - 2 {M} r^2.
\ee
The motivation for writing the metric in this form is that it then
resembles the higher dimensional anti-de Sitter Kerr solutions. We
have chosen the normalisation of the time and angular coordinates so
that the latter has the usual period and the former has norm $r l$ at
spatial infinity. Rewriting the BTZ black hole metric in Kerr-Schild
and Boyer-Lindquist type coordinates was discussed
recently in \cite{Kim} in the context of studying the global structure 
of the black hole. Using the coordinate transformations 
\ba
T = t; && \hspace{10mm} \Phi = \phi + a l^2 t; \\
R^2 &=& \frac{1}{\Xi}(r^2 + a^2),
\ea
with $\Xi = 1 - a^2/l^2$, followed by a shifting of the radial
coordinate, we can bring the metric back into the usual BTZ form. 
The horizons are defined by the zero points of $\Delta_r$, with the
event horizon being at 
\be
r_{+}^2 = \frac{1}{2 l^2}(2 {M} - 1 - a^2 l^2) + \frac{1}{2 l^2} \sqrt{
(1 + a^2 l^2 - 2 {M})^2 - 4 a^2 l^2}.
\ee
Expressed in terms of the variables $({M},a)$ the supersymmetry and
cosmic censorship conditions become 
\be
{M} \ge \frac{1}{2} (1 + \left | a \right | l)^2, \label{up_ab}
\ee
where the choice of sign of $a$ determines which Killing spinor is
conserved in the BPS limit. In the special case $\bar{M} \equiv 0$ 
both supersymmetries are
preserved; this is true for all $a$ and not just for the limiting value
$\left | a \right | l \rightarrow  1$ which saturates (\ref{up_ab}).

\bigskip

As is the case in higher dimensions, the ${M}=0$ metric is identified
three-dimensional anti-de Sitter space. One can calculate the inverse
temperature of the black hole to be 
\be
\beta_{t} = 4 \pi \frac{r_{+}^2 + a^2}{\Delta_{r}'(r_{+})}.
\ee
In the calculation of the action, only the volume term
contributes; the appropriate background is the ${M}=0$ solution with the
imaginary time coordinate scaled so that the geometry matches on a
hypersurface of large radius 
\be 
\tau \rightarrow ( 1 - \frac{M}{l^2 R^2}) \tau.
\ee
Then the action is given by 
\be 
I_3 = - \frac{\pi (r_{+}^2 + a^2)}{\Xi \Delta_{r}'(r_{+})}  
\left [ r_{+}^2 l^2 + a^2 l^2 - {M} \right ].
\ee
In this coordinate system the thermodynamic quantities can be written
as
\ba
{\cal M}' = \frac{M}{4 \Xi}; & \hspace{5mm} & J' = \frac{M a }{2\Xi^2};
\\
\Omega' = \frac{\Xi a }{(r_{+}^2 + a^2)}; & \hspace{5mm} & S =
\frac{\pi}{2 \Xi r_{+}} (r_{+}^2 + a^2). \nonumber 
\ea
We now have to decide how to take the limit of critical angular
velocity in this coordinate system. 
The key point is that this coordinate system is not adapted
to the rotating Einstein universe on the boundary. The angular
velocity of the black hole in this coordinate system 
vanishes in the limit $al \rightarrow 1$ and is always smaller
in magnitude than $l$. 

In both this and following sections, we shall
adhere to the notation that primed thermodynamic quantities are expressed with
respect to the Kerr coordinate system whilst unprimed thermodynamic 
quantities are expressed with respect to the Einstein universe
coordinate system. We also assume from here onwards that $a$ is positive. 

\bigskip

The angular velocity of the rotating Einstein universe is given by 
\be
\Omega = \Omega' + al^2; \label{om2}
\ee 
that is, we need to define the angular velocity with respect to 
the coordinates $(T, \Phi)$. Now suppose that $\Omega  =
 l (1 - \epsilon)$ where $\epsilon$ is small. This requires that 
\be
\epsilon = (1 - al) \frac{(r_{+}^2 - a/l)}{(r_{+}^2 + a^2)}.
\ee
For the Einstein universe on the boundary to be rotating at the
critical angular velocity, either $al =1$ or $r_{+}^2 = a/l$. 
Note that not only the action but also 
the entropy is divergent in the limit $al =1$.  

Let us explore the limit $r_{+}^2 = a/l$ first;
it is straightforward to show that this coincides with the
supersymmetry limit. This means that in every supersymmetric black
hole the boundary is effectively rotating at the speed of light, which
is apparent from the limit of $\Omega$ given in (\ref{mass_btz}).
Cosmic censorship requires that $r_{+}^2 \ge a/l$ and hence 
the rotating Einstein universe never rotates faster than the speed of
light. Put another way, any BTZ black hole can be in equilibrium with
thermal radiation in infinite space, no matter what its mass is. 

The metric of a supersymmetric BTZ black hole is 
\be
ds^2 = - l^2 y^2 dT^2 + \frac{jl}{2} ( dT - l^{-1} d\Phi)^2 + y^2
d\phi^2 + \frac{dy^2}{l^2 y^2}.
\ee
Now starting from the black hole metric (\ref{alt_met}) and using
the coordinate transformations 
\ba
T = t; \hspace{10mm} \Phi = \phi + al^2 t; \\
y^2 = \frac{1}{\Xi} (r^2 - a/l), \nonumber
\ea
the general supersymmetric metric can also be expressed as
\be
ds^2 = - l^2 y^2 dT^2 + y^2 d\Phi^2 + \frac{dy^2}{l^2 y^2} +
\frac{al ( 1 + al)^2}{\Xi^2} ( dT - l^{-1} d\Phi)^2.
\ee
Correspondence between the two metrics requires that   
\be 
m = j l =  \frac{2 a l}{(1-al)^2}.
\ee
So a supersymmetric black hole has a mass which diverges as we take
the limit $al \rightarrow 1$. This is apparent if we define the
thermodynamic quantities with respect to the
coordinates $(T, \Phi)$. The energy and inverse temperature are
unchanged (${\cal M} \equiv {\cal M}'$ and $\beta_t \equiv \beta$) whilst 
\be
J = \frac{Ma}{2 \Xi (1 + l^2 r_{+}^2)}.
\ee 
So the mass and angular momentum of {\it any} black hole diverge as we take
the limit $al \rightarrow 1$. 
It is useful to consider (very non-extreme) black holes which are at high
temperature; this requires that $r_{+} l \gg 1$ and so if we define a
dimensionless inverse temperature 
\be
\bar{\beta} = l \beta \approx \frac{2 \pi}{l r_{+}},
\ee
we find that the other thermodynamic quantities behave for $al
\rightarrow 1$ as 
\ba
I_{3} = - \frac{\pi^2}{l \Xi \bar{\beta}}; & \hspace{5mm} & 
S = \frac{\pi^2}{l \Xi \bar{\beta}} \\
{\cal M} = \frac{\pi^2}{2 l \Xi \bar{\beta}^2}; & \hspace{5mm} & 
J = \frac{1}{4 l^2 \Xi}, \nonumber
\ea
where the latter two quantities are defined with respect to the
dimensionless temperature. These thermodynamic quantities are
consistent both with the thermodynamic relations, and with the result
for the partition function of the corresponding conformal field
theory. 

\section{Rotating black holes in four dimensions}
\label{sec:Four}

Rotating black holes in four dimensions with asymptotic AdS behaviour
were first constructed by Carter \cite{Carter68} many years ago. 
There has been interest in such solutions recently as solitons
of $N=2$ gauged supergravity in four dimensions
\cite{KosteleckyPerry95} and in the   
context of topological black holes \cite{CaldarelliKlemm98}.  The metric is
\be
 ds^2 =-\frac{\Delta_r}{\rho^2}
      \left[dt - \frac{a}{\Xi} \sin^2\theta d\phi\right]^2
  + \frac{\rho^2}{\Delta_r}dr^2 + \frac{\rho^2}{\Delta_\theta}d\theta^2
  + \frac{\sin^2\theta \Delta_\theta}{\rho^2}\left[adt - 
      \frac{(r^2+a^2)}{\Xi} d\phi \right]^2
\ee
where
\ba
 \rho^2 & = & r^2 + a^2\cos^2\theta \nonumber \\
 \Delta_r & = & (r^2 + a^2)(1+ l^2 r^2) - 2Mr \\
 \Delta_\theta & = & 1 - l^2 a^2 \cos^2\theta \nonumber \\
 \Xi & = & 1 - l^2 a^2 \nonumber 
\ea
The parameter $M$ is related to the mass, $a$ to the angular momentum and
$l^2 = -\Lambda/3$ where $\Lambda$ is the (negative) cosmological 
constant.  The solution is valid for $a^2<l^2$, but becomes singular
in the limit $a^2=l^2$ which is the focus of our attention here.   
The event horizon is located at $r=r_+$, the largest root of the
polynomial $\Delta_r$.  One can define a critical mass parameter
$M_{e}$ such that \cite{CaldarelliKlemm98}
\begin{equation}
M_{e} l = \frac{1}{3 \sqrt{6}} \left ( \sqrt{ (1+a^2 l^2)^2 + 12 a^2 l^2 } + 2
(1+a^2 l^2) \right ) \left (\sqrt{ (1+a^2 l^2)^2 + 12 a^2 l^2 } -
(1+a^2 l^2) \right )^{\frac{1}{2}}. \nonumber 
\end{equation}
Cosmic censorship requires that $M \ge M_{e}$ with the limiting case
representing an extreme black hole. In the limit of critical angular
velocity, the bound becomes
\be
M l \ge \frac{8 }{3 \sqrt{3}}, \label{crit_mas}
\ee
which we will see implies that physical black holes must be at least
as large as the cosmological scale. The angular velocity $\Omega'$ is 
\be
 \Omega' = \frac{\Xi a}{(r_+^2 + a^2)}, 
\ee
whilst the area of the horizon is 
\be
 {\cal A} = 4\pi \frac{r_+^2 + a^2}{\Xi},
\ee
and the inverse temperature is 
\be
 \beta_{t} = \frac{4\pi (r_+^2 + a^2)}{\Gamma_r'(r_+)} = 
    \frac{4\pi (r_+^2 + a^2)}{r_{+}(3l^2 r_+^2 + (1+ a^2 l^2)
  - a^2/r_+^2)}.
\ee
We should mention here the issue of the normalisation of the
Killing vectors and the rescaling of the associated coordinates. 
One choice of normalisation of the Killing vectors ensures that
the associated conserved quantities generate the $SO(3,2)$ algebra:
this was the natural choice in the context of
\cite{KosteleckyPerry95}. Here we have chosen the metric so that the
coordinate $\phi$ has the usual periodicity whilst the norm of the
imaginary time Killing vector at infinity is $l r$. Note that we are
referring to the issue of the normalisation of the Kerr coordinates rather
than to the relative shifts between Kerr and Einstein universe
coordinates.  

\bigskip

If we Wick rotate both the time coordinate 
and the angular momentum parameter, 
\be
 t = -i\tau \spaceand a = i\alpha, 
\ee
then we obtain a real Euclidean metric where the radial coordinate
is greater than the largest root of $\Delta_{r}$. 
The surface $r=r_+$ is a bolt of the co-rotating Killing vector,
$\xi = \partial_\tau + i \Omega\partial_\phi$.
However, an identification of imaginary time coordinates 
must also include a rotation through $i \beta \Omega$ 
in $\phi$; that is, we identify the points
\be
 (\tau, r, \theta, \phi) \sim (\tau + \beta, r, \theta, \phi + i \beta\Omega).
\ee
We now want to calculate the Euclidean action, defined as 
\be
I_4 = - \frac{1}{16 \pi} \int d^4x \sqrt{g} \left [ R_g + 6 l ^2
\right ],
\ee 
where we have set the gravitational constant to one. The choice of
background is made 
by noting that the $M=0$ Kerr-AdS metric is actually the AdS metric in
non-standard coordinates \cite{HenneauxTeitelboim85}.  If we make the 
implicit coordinate transformations
\ba
 T  =  t  & \hspace{5mm} &  
 \Phi  =  \phi - a l^2 t \nonumber \\
 y \cos{\Theta} & = & r \cos\theta \\
 y^2 & = & \frac{1}{\Xi} [ r^2 \Delta_\theta + a^2 \sin^2\theta ]
\nonumber 
\ea  
this takes the AdS metric,
\be
 d\tilde{s}^2 = - (1+ l^2 y^2) dT^2 + 
    \frac{1}{1 + l^2 y^2} dy^2 + 
 y^2(d\Theta^2 + \sin^2\Theta d\Phi^2),
\ee
to the $M=0$ Kerr-AdS form. To calculate the action we need to match
the induced Euclidean 
metrics on a hypersurface of constant radius $R$ by scaling 
the background time coordinate as
\be
 \tau \rightarrow \left(1 - \frac{M}{l^2R^3} \right) \tau, 
\ee
and then we find that 
\be
 I_4 = - \frac{\pi (r_+^2 + a^2)^2 (l^2r_+^2 - 1)}
          {\Xi r_{+} \Delta_r'(r_+)} 
 =  - \frac{\pi (r_+^2 + a^2)^2 (l^2r_+^2 - 1)}
            {(1-l^2 a^2) (3l^2r_+^4 + (1+l^2a^2)r_{+}^2 - a^2)}. 
\ee 
Features of this result are as follows. 
The action is positive for $r_{+}^2 \le 1/l^2$ and negative for larger
$r_{+}$; just as for Schwarzschild anti-de Sitter this indicates
that there is a phase transition as one increases the mass. The action
is clearly divergent for extreme black holes as one would
expect. There is also a divergence when $\Xi \rightarrow 0$; for small
radius black holes the action diverges to positive infinity, whilst
for large radius black holes the action diverges to negative
infinity. In the special case $r_{+}^2 = a/l$ the action is finite and
positive in the limit $al \rightarrow 1$. 

\bigskip

Defining the mass and the angular momentum of the black hole as
\be 
{\cal M}' = \frac{1}{8 \pi} \int \nabla_{a} \delta {\cal T}_{b} dS^{ab}
\hspace{5mm} J' = \frac{1}{4\pi} \int \nabla_{a} \delta {\cal J}_{b}
dS^{ab},
\ee
where ${\cal T}$ and ${\cal J}$ are the generators of time translation and
rotation respectively and one integrates the difference between the
generators in the spacetime and the background over a celestial sphere
at infinity, then we find that
\be
{\cal M}' = \frac{M}{\Xi}; \hspace{5mm} J' = \frac{a M}{\Xi^2}. 
\ee
Allowing for the differences in normalisation of the generators, these
values agree with those given in \cite{KosteleckyPerry95}.
Using the usual thermodynamic relations we can check that the entropy
is 
\be
 S = \pi \frac{r_+^2+a^2}{\Xi},
\ee
as expected. Note that none of the extreme black holes are
supersymmetric: in four dimensions there needs to be a non vanishing
electric charge for such black holes, regarded as solutions of a
gauged supergravity theory, to be supersymmetric. 

\bigskip

It is well known that small Schwarzschild anti-de Sitter black holes are
thermodynamically unstable in the sense that their heat capacity is
negative, just as for Schwarzschild black holes in flat space. We find
such an instability in dimensions $d \ge 4$ for black holes whose
radius is less than a critical radius which is dimension dependent but
is approximately $1/l$. One can show that small rotating black holes
are also unstable in this sense but only for rotation parameters of
order $0.1 l^{-1}$ or less (again the precise limit is dimension
dependent); larger angular velocities stabilise the black holes. 
In three dimensions no anti-de Sitter black holes have negative
specific heat. 

\bigskip

\begin{figure}
\begin{center}
\leavevmode
\bigskip
\epsfxsize=.45\textwidth
\epsfbox{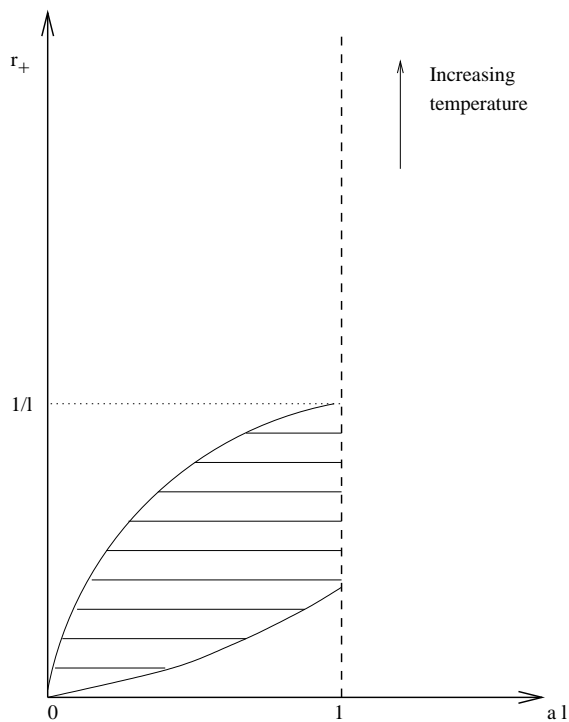}
\bigskip
\caption[]
{Plot of black hole radius $r_{+}$ against $al$. For $r_{+} < 1/l$,
  the action is positive, whilst the action blows up along the line
  $al = 1$. The lower line denotes the radius of the extreme black hole
  $r_{c}$ as a function of $a$.  
  In the hatched region $r_{c}^2 \le r_{+}^2 < a/l$ the
  Einstein universe on the boundary rotates faster than the speed of
  light. The action is finite and positive at $r_{+}^2 = a/l$ but infinite
  and positive for extreme black holes. In three dimensions the
  supersymmetric limit coincides with $r_{+}^2 = a/l$, whilst in five
  and higher dimensions the cosmological bound is at $r_c = 0$.  }
\label{fig:fig0}
\end{center}
\end{figure}

To take the limit of critical angular velocity, we need to use the
coordinate system adapted to the rotating Einstein universe. As in 
three dimensions the angular velocity of the Einstein
universe is given by 
\be
\Omega = \Omega' + al^2,
\ee
and is defined with respect to the coordinates $(T, \Phi)$. Defining
$\Omega = l (1 -\epsilon)$ as before we find that 
\be 
\epsilon = ( 1 - al) \frac{(r_{+}^2 -a/l)}{(r_{+}^2 + a^2)}. \label{eps}
\ee
Rotation at the critical angular velocity hence requires that either
$al = 1$ or $ r_{+}^2 = a/l$, as in three dimensions.
Generically the thermodynamics of the
four dimensional black hole are similar to those of the BTZ black
hole, and in fact to those of higher dimensional black holes also. 
The $(r_{+}, a)$ plane for a single parameter black hole in a general 
dimension is illustrated in Figure~\ref{fig:fig0}.

\bigskip

There is however a novelty compared to the
three dimensional case. The cosmological bound permits solutions with
$r_{+}^2 < a/l$; for example, in the limiting case $al =1$, the
extreme solution has $r_{+}^2 = a^2/3$. To preserve the Lorentzian
signature of the metric we require that $al \le 1$, and so $\Omega' >
l$ in the limit $r_{+}^2 < a/l$. That is, only for sufficiently
large black holes can one have the rotating black hole in
equilibrium with thermal radiation in infinite space. This is
reflected in the fact that the action changes sign at $r_{+} =
1/l$. In the limit of zero curvature - by taking $l$ to zero - we 
find, as expected, that there are no rotating 
black holes for which there is a Killing vector which is timelike
right out to infinity. 

One can rewrite the thermodynamic quantities of the black hole with
respect to the coordinate system $(T, \Phi)$. The temperature is
unchanged ($\beta \equiv \beta_t$) whilst the energy and angular
momentum are given by 
\be
{\cal M} = {\cal M}'; \hspace{5mm} J = \frac{a M}{ \Xi (1+ l^2 r_{+}^2)}.
\ee
We are particularly interested in the limit of the action as $al
\rightarrow 1$ at high temperature. Defining a dimensionless quantity
\be
\hat{\beta} = l \beta \approx \frac{4 \pi}{3 l r_{+}},
\ee
where the latter relation applies in the high temperature limit, the
action diverges as
\be
I_{4} = - \frac{8 \pi^3}{27 l^2 \bar{\beta}^2 (1-al)} 
\ee
The other thermodynamic quantities behave to leading order as 
\ba 
(1 - \Omega) = (1 - al); & \hspace{5mm} &
{\cal M} = \frac{16 \pi^3}{27 l^2 \bar{\beta}^3 (1-al)} \\
J = \frac{\pi}{3 l^3 \bar{\beta} (1-al)} & \hspace{5mm} &
S = \frac{8 \pi^3}{9 l^2 \bar{\beta}^2 (1-al)} \nonumber 
\ea
The entropy diverges at the critical value, as do the energy and the
angular momentum.  Note that the divergence of the angular momentum is
subleading in $\bar{\beta}$.
As we stated in the introduction, there is no sense in which one can
take the energy to be finite in the critical limit. If we take
${\cal{M}}$ to be fixed, then $M$ must approach zero in the limit.
However according to (\ref{crit_mas}) there is no horizon unless the
mass parameter $M$ is of the cosmological scale.

\section{Rotating black holes in five dimensions}
\subsection{Single parameter anti-de Sitter Kerr black holes}

We now consider rotating black holes within a five dimensional anti-de
Sitter background. In five dimensions the rotation group is $SO(4)
\cong SU(2)_{L} \times SU(2)_{R}$. Black holes may be characterised by
two independent projections of the angular momentum vector which may
be denoted as the angular momenta $J_{L}$ and $J_{R}$. This is the
most natural parametrisation when one considers the conformal field
theory describing such states but the usual construction of Kerr
metrics in higher dimensions will use instead two parameters
$J_{\phi}$ and $J_{\psi}$ which we choose such that
\be
J_{L,R} = \left ( J_{\phi} \pm J_{\psi} \right ),
\ee
where we express the metric on the three sphere in the form 
\be
ds^2 = d\theta^2 + \sin^2\theta d\phi^2 + \cos^2\theta d\psi^2. 
\ee
The two classes of special cases may be represented by the limits
\ba
J_{R} = 0 \hspace{2mm} & \Rightarrow & \hspace{2mm} J_{\phi} = J_{\psi};
\\
J_{L} = J_{R} \hspace{2mm} & \Rightarrow & \hspace{2mm} J_{\psi} = 0.
\ea
The former case will be considered in the next subsection.
As for the stationary asymptotically flat solutions constructed by
Myers and Perry \cite{My_Pe}, the single parameter 
Kerr anti-de Sitter solution in
$d$ dimensions follows straightforwardly from the four-dimensional
solution. It is convenient to write it in the form 
\ba
 ds^2 & = & - \frac{\Delta_r}{\rho^2}(dt - \frac{a}{\Xi}
  \sin^2\theta d\phi)^2 + 
     \frac{\rho^2}{\Delta_r}dr^2 + \frac{\rho^2}{\Delta_\theta}d\theta^2
    \nonumber \\
   & & + \frac{\Delta_\theta\sin^2\theta}{\rho^2}[adt -
\frac{(r^2+a^2)}{\Xi} d\phi]^2
    + r^2\cos^2\theta d\Omega_{d-4}^2
\ea
where $d\Omega_{d-4}^2$ is the unit round metric on the $(d-4)$ sphere and 
\ba
 \Delta_r & = & (r^2+a^2)(1+l^2r^2) - 2M r^{5-d}; \nonumber \\
 \Delta_\theta & = & 1 - a^2 l^2 \cos^2\theta; \\
 \Xi & = & 1 - a^2 l^2; \nonumber \\ 
 \rho^2 & = & r^2 + a^2\cos^2\theta. \nonumber 
\ea
The angular velocity on the horizon is all dimensions is
\begin{equation}
\Omega' = \frac{\Xi a}{(r_{+}^2 + a^2)}.
\end{equation}
The thermodynamics of single parameter solutions are generically
similar in all dimensions. In five dimensions 
we can solve explicitly for the horizon position finding that 
\begin{equation}
r_{+}^2 = \frac{1}{2l^2} [ \sqrt{(1- a^2 l^2)^2 + 8 M l^2} -
(1-a^2 l^2) ].
\end{equation}
The condition for a horizon to exist
is that $r_{+}$ must be real, which requires that $2 M \ge a^2$. 
The volume of the horizon is 
\begin{equation}
V = \frac{2 \pi^2}{\Xi} r_{+} (r_{+}^2 + a^2),
\end{equation}
and the inverse temperature is
\begin{equation}
\beta_{t} = 4 \pi \frac{(r_{+}^2 + a^2)}{\Delta_{r}'(r_{+})} 
= \frac{2 \pi (r_{+}^2 + a^2)}{r_{+} (2 l^2 r_{+}^2 + 1 +
a^2 l^2)}.
\end{equation}
It is useful to note that the $M=0$ Kerr anti-de Sitter solution
reduces to the anti-de Sitter background, with points identified in
the angular directions, for all $d$: this follows
from the same coordinate transformation as for the four dimensional
solution. The same coordinate transformation also brings 
the $M \neq 0$ solution into a manifestly asymptotically anti-de Sitter form. 

In calculating the action the appropriate background
is the $M=0$ solution, with the imaginary time coordinate rescaled so
that the induced metric on a hypersurface of large radius $R$ matches
that of the $M \neq 0$ solution. This requires that we scale
\begin{equation}
\tau \rightarrow (1 - \frac{M}{R^4 l^2}) \tau.
\end{equation}
The volume term in the action is given by    
\begin{equation}
I_{5} = - \frac{1}{16 \pi} \int d^5x (R_g + 12 l^2), \label{5d_act}
\end{equation}
(with the gravitational constant equal to one) and 
the surface term does not contribute. Evaluating the volume term we find
that the action is given by 
\begin{equation}
I_{5} =  \frac{\pi^2}{4 \Xi} \frac{(r_{+}^2 + a^2)^2 (1-
l^2 r_{+}^2)}{r_{+} (2l^2 r_{+}^2  + 1 + a^2 l^2)}.
\end{equation}
This action has the same generic features as in the lower dimensional 
cases, namely (i) the sign changes at the critical radius $r_{+}^2 =
1/l^2$; (ii) the action diverges as $\Xi \rightarrow 0$ except for black
holes of the critical radius $r_{+}^2 = a/l$.

It is straightforward to show that the mass and angular momentum of
the rotating black hole with respect to the anti-de Sitter background
are given by
\be
{\cal M}' = \frac{3 \pi M}{4 \Xi}, \hspace{5mm} J_{\phi}' = \frac{\pi M a
}{2 \Xi^2}.
\ee
Then the usual thermodynamic relations give the entropy of
the black hole as
\be
S = \beta ({\cal M} + \Omega J) - I_{5} = \pi^2 \frac{r_{+} (r_{+}^2 +
a^2)}{2 \Xi},
\ee
which is related to the horizon volume in the expected way. 

\bigskip

It is interesting to note that both the temperature and the entropy
vanish for black holes with horizons at $r_{+} = 0$, even though the
mass and angular momentum can be non-zero. Since a necessary (though
non-sufficient) condition for a black hole to be supersymmetric is
that the temperature vanishes, only states for which the bound $2 M =
a^2$ is saturated could  be supersymmetric. 

Just as the four-dimensional rotating black holes are solutions of
$N=2$ gauged supergravity in four dimensions, so we can regard the
five-dimensional solutions as solutions of a five
dimensional gauged supergravity theory. However, as in the four
dimensional case, the black holes do not preserve any supersymmetry
for non-zero mass unless they are charged. 

One can see this as follows. Supersymmetry requires the existence of a
supercovariantly constant spinor $\epsilon$ satisfying 
\be 
\delta \Psi_{m} = \hat{D}_{m} \epsilon = (\nabla_{m} +
\frac{1}{2} i l \gamma_{m}) \epsilon = 0,
\ee
where $\Psi$ is the gravitino, $\hat{D}$ is the supercovariant
derivative, $\nabla$ is the covariant derivative and $\gamma$ is a
five-dimensional gamma matrix. The integrability condition then
becomes
\be
\left [ \hat{D}_{m}, \hat{D}_{n} \right ] \epsilon = 0 \Rightarrow
\left (R_{mnab} \gamma^{ab} + 2 l^2 \gamma_{mn} \right ) \epsilon = 0,
\label{int_con}
\ee
where $a,b$ are tangent space indices. It is straightforward to verify
that all components of the bracketed expression 
vanish for the background whilst
for the rotating black hole the integrability conditions reduce to 
\be 
\frac{M}{\Xi} \gamma_{a} \epsilon = 0.
\ee
Hence only in the zero mass black hole - anti de Sitter space itself -
is any supersymmetry preserved. We expect that supersymmetry can be
preserved if we include charges, but leave this as an
issue to be explored elsewhere \cite{mmt}. General static charged
solutions of $N=2$ gauged supergravity in five dimensions have been
discussed recently in \cite{BCS}; one can construct the natural
generalisations to general stationary black holes starting from the 
neutral five dimensional stationary solutions given here. One can also
construct solutions for which the horizon is hyperbolical rather than
spherical; such solutions are analogous to those discussed in 
\cite{CaldarelliKlemm98}.

\bigskip

Taking the limit of critical angular velocity requires that we
move to the coordinates $(T,\Phi)$ which are adapted to the rotating
Einstein universe. Then letting $\Omega = l (1- \epsilon)$ with
$\epsilon$ defined as in (\ref{eps}) we find that in the critical limit 
either $r_{+}^2 = a/l$ or $al = 1$. Since cosmic censorship requires
that $r_{+} \ge 0$ with equality in the extreme limit, we can again
have solutions for which $\Omega > l$ which in turn implies that the
black holes cannot be in equilibrium with radiation right out to
infinity. As in four dimensions the action changes sign at the
critical value $r_{+}^2 = a/l$. 

The thermodynamic quantities relative to the coordinate system $(T,
\Phi)$ are $\beta \equiv \beta_t$, ${\cal M} \equiv {\cal M}'$ and 
\be 
J_{\Phi} = \frac{\pi M a }{2 \Xi ( 1 + l^2 r_{+}^2)}.
\ee
In the limit $al \rightarrow 1$ at high temperature such that  
\be
\bar{\beta} = l \beta \approx \frac{\pi}{ l r_{+}},  
\ee
we can express the thermodynamic quantities as 
\ba 
I_{5} \approx  - \frac{\pi^5}{8 l^3 \Xi \bar{\beta}^3}; & \hspace{5mm}
& {\cal M} \approx  \frac{3 \pi^5}{8 l^3 \Xi \bar{\beta}^4};
\label{5d_1a} \\ 
J_{\Phi}  \approx  \frac{\pi^3}{2 l^4 \Xi \bar{\beta}^2}; &
\hspace{5mm} &  S  \approx  \frac{\pi^5}{2 l^3 \Xi \bar{\beta}^3}
\nonumber 
\ea
where the energy and angular momentum are defined with respect to the
dimensionless temperature. Note that the angular momentum is again 
subleading in $\bar{\beta}$ dependence relative to the mass and the action. 
The divergence at critical angular velocity is in agreement with that
of the conformal field theory. 

\subsection{General five-dimensional AdS-Kerr solution}

The metric for the two parameter five-dimensional rotating black hole
is given by 
\ba
ds^2 &=& - \frac{\Delta}{\rho^2} (dt - \frac{a \sin^2\theta}{\Xi_a}d\phi -
\frac{b \cos^2\theta}{\Xi_b} d\psi)^2 +
\frac{\Delta_{\theta}\sin^2\theta}{\rho^2} (a dt -
\frac{(r^2+a^2)}{\Xi_a} d\phi)^2 \nonumber \\
&& + \frac{\Delta_{\theta}\cos^2\theta}{\rho^2} (b dt -
\frac{(r^2+b^2)}{\Xi_b} d\psi)^2 + \frac{\rho^2}{\Delta} dr^2 +
\frac{\rho^2}{\Delta_{\theta}} d\theta^2 \\
&& + \frac{(1+r^2 l^2)}{r^2 \rho^2}
\left ( ab dt - \frac{b (r^2+a^2) \sin^2\theta}{\Xi_a}d\phi - \frac{a
    (r^2 + b^2) \cos^2 \theta}{\Xi_b} d\psi \right )^2, \nonumber
\ea
where 
\ba
\Delta &=& \frac{1}{r^2} (r^2 + a^2) (r^2 + b^2) (1 + r^2 l^2) - 2M;
\nonumber \\
\Delta_{\theta} &=& \left ( 1 - a^2 l^2 \cos^2\theta - b^2 l^2
  \sin^2\theta \right ); \\
\rho^2 &=& \left ( r^2 + a^2 \cos^2\theta + b^2 \sin^2\theta \right);
\nonumber \\
\Xi_a &=& (1-a^2 l^2); \hspace{5mm} \Xi_b = (1 -b^2 l^2). \nonumber 
\ea
It should be straightforward to construct the metric for 
general rotating black holes in anti-de Sitter backgrounds of
higher dimension.  
As for the single parameter solution, the $M=0$ metric is anti-de
Sitter space, with points identified in the angular directions. 
Using the coordinate transformations 
\ba
\Xi_a y^2 \sin^2\Theta &=& (r^2 + a^2) \sin^2\theta; \nonumber \\
\Xi_b y^2 \cos^2\Theta &=& (r^2 + b^2) \cos^2\theta; \nonumber
\\
\Phi &=& \phi + a l^2 t; \\
\Psi &=& \psi + b l^2 t, \nonumber 
\ea
the metric can be brought into a form which is manifestly asymptotic
to anti-de Sitter spacetime. The parameters $a$ and $b$ are
constrained such that $a^2, b^2 \le l^{-2}$ and the metric is only
singular if either or both parameters saturate this limit. 

Defining the action as in (\ref{5d_act}) we find that 
\be
I_5 = - \frac{\pi \beta l^2}{4 \Xi_a \Xi_b} \left [ (r_{+}^2 + a^2)(r_{+}^2
  + b^2) - M l^{-2} \right ],
\ee
where the inverse temperature is given by
\be
\beta_t = 
\frac{4 \pi (r_{+}^2 + a^2) (r_{+}^2 + b^2)}{ r_{+}^2 \Delta'(r_{+})},
\ee
and $r_{+}$ is the location of the horizon defined by 
\be
(r_{+}^2 + a^2)(r_{+}^2 + b^2) (1 + r_{+}^2 l^2) = 2 M r_{+}^2.
\ee
For real $a$, $b$ and $l$ there are two real roots to this equation;
when $a = b$ these coincide to give an extreme black hole when 
\ba
r_c^2 &=& = \frac{1}{4 l^2} \left ( \sqrt{1 + 8 a^2 l^2} - 1 \right );
\\
2 M_c l^2 &=&  \frac{1}{16} \left ( \sqrt{1 + 8 a^2 l^2} - 1 + 4 a^2 l^2
\right ) \left ( 3 \sqrt{1 + 8 a^2 l^2} + 5 + 4 a^2 l^2 \right). \nonumber
\ea
The entropy of the general two parameter black hole is given by 
\be
S = \frac{\pi^2}{2 r_{+} \Xi_a \Xi_{b}} (r_{+}^2 + a^2)(r_{+}^2 + b^2),
\ee
whilst the mass and angular momenta are 
\be
{\cal M}' = \frac{3 \pi M}{4 \Xi_a \Xi_b}; \hspace{2.5mm}
J_{\phi}' = \frac{\pi M a }{2 \Xi_a^2}; \hspace{2.5mm} \hspace{2.5mm}
J_{\psi}' = \frac{\pi M b }{2 \Xi_b^2}, \nonumber 
\ee
with the angular velocities on the horizon being
\be
\Omega_{\phi}' = \Xi_a \frac{a}{r_{+}^2 + a^2}; \hspace{5mm}
\Omega_{\psi}' = \Xi_b \frac{b}{r_{+}^2 + b^2}. 
\ee
Since the black hole is singular only when either or both of $\Xi_a$  
and $\Xi_b$ tend to zero, we should look in particular at the latter
case for which the two rotation parameters $a$ and $b$ are equal in
magnitude. Then we can write the metric in the transformed coordinates
as
\ba
ds^2 &=& - (1 + y^2 l^2) dT^2 + y^2 \left (
  d\Theta^2 + \sin^2\Theta
  d\Phi^2 + \cos^2\Theta d\Psi^2 \right) \nonumber \\
&& + \frac{2M}{y^2 \Xi^2} ( dT - a \sin^2 \Theta
d\Phi - a \cos^2\Theta d\Psi)^2 + \frac{y^4 
dy^2}{( y^4(1+ y^2l^2) -
  \frac{2M}{\Xi^2}y^2 + \frac{2 M a^2}{\Xi^3})},  \label{a=b}
\ea
where $\Xi = 1 - a^2 l^2$. The position of the horizon of the extreme
solution in these coordinates is 
\be
y^2 = \frac{1}{4 \Xi} \left [ 4 a^2 l^2 - 1 + \sqrt{1 + 8 a^2 l^2}
\right ].
\ee
In the critical limit, $al \rightarrow 1$, the size
of the black hole becomes infinite in this coordinate system. 

One can check to see whether the integrability condition
(\ref{int_con}) is satisfied by the black hole metric
(\ref{a=b}). Preservation of supersymmetry requires that 
\be
\frac{M}{\Xi^2} \gamma_a \epsilon = 0,
\ee
and hence only in the zero mass black hole is any supersymmetry
preserved. We have not checked the integrability condition in the
general two parameter rotating black hole but do not expect
supersymmetry to be preserved. In three dimensions the integrability
condition is trivially satisfied since the BTZ black hole is locally
anti-de Sitter and supersymmetry preservation relates to global
effects. In higher dimensions it does not seem possible to satisfy the
integrability conditions without including gauge fields. 

\bigskip

The Einstein universe rotates at the speed of light in at least some
directions either when one or both of $\Xi_a$ and $\Xi_b$ vanish or
when $r_{+}^2 =a/l$ or when $r_{+}^2 = b/l$. 
The action is singular when either or both of $\Xi_a$ and $\Xi_b$ are
zero and when the black hole is extreme. The action is positive for 
$r_{+} \le 1/l$; there is a phase transition as the mass of the 
black hole increases. 

If $r_{+}^2 = a/l$ the action
will be positive and finite when $\Xi_a$ vanishes and positive and
infinite when $\Xi_b$ vanishes. For $l r_{+}^2 < {\rm Max} \left [ a,b
\right ]$ there will be directions in the Einstein universe which are
rotating faster than the speed of light. In the limiting case $a=b$ the action
diverges for all $r_{+}$ as $\Xi$ tends to zero. 

In the high temperature limit, the action for the equal parameter
rotating black hole diverges as  
\be 
I_5 = - \frac{\pi^5} {8 l^3 \Xi^2 \bar{\beta}^3},
\ee
with $\bar{\beta} \approx \pi/(r_{+} l) \ll 1$. The other
thermodynamic quantities follow easily from this expression, and are
in agreement with those derived from the conformal field theory in
section two.  

We should mention what we expect to happen in higher dimensions. A
generic rotating black hole in $d$ dimensions will be classified by 
$\left [ (d-1)/2 \right ]$ rotation parameters $a_{i}$, where $[x]$
denotes the integer part of $x$. Thus we expect both the action and the
metric to be singular if any of the $a_{i}$ vanish. Provided that the
black hole horizon is at $r_{+} > 1/l$ the action should diverge to
negative infinity in the critical limit, behaving as 
\be
I_{d} \sim - \frac{1}{\beta^{d-2} \prod_{i} \epsilon_{i}},
\ee
where $\epsilon_{i} = 1 - \Omega_{i}$ and the product is taken over
all $i$ such that $a_{i}l \rightarrow 1$. The $\beta$ dependence
follows from conformal invariance, whilst one should be able to derive
the $\epsilon_i$ dependence by looking at the behaviour of the spherical
harmonics.

\end{document}